\documentclass[prd,aps,preprint,amsmath,amssymb,nofootinbib]{revtex4-1}
\usepackage{verbatim,graphics,graphicx,color,slashed,textcomp}
\usepackage[normalem]{ulem} 
\usepackage[colorlinks=true,linkcolor=red,urlcolor=blue,citecolor=blue]{hyperref}

\graphicspath{{figures/}}

\usepackage[utf8]{inputenc}

\newcommand{\keV}{\mathrm{keV}}
\newcommand{\MeV}{\mathrm{MeV}}
\newcommand{\GeV}{\mathrm{GeV}}

\begin{document}
	
\title{Semi-annihilating $Z_3$ Dark Matter for XENON1T Excess}
	\author{P. Ko${}^{(a)}$ and Yong Tang${}^{(b,c,d,e)}$}
	\affiliation{\begin{footnotesize}
		${}^a$Korea Institute for Advanced Study, Seoul 02445, South Korea\\
		${}^b$School of Astronomy and Space Sciences, University of Chinese Academy of Sciences (UCAS), Beijing 100049, China\\
		${}^c$School of Fundamental Physics and Mathematical Sciences, \\
		Hangzhou Institute for Advanced Study, UCAS, Hangzhou 310024, China \\
		${}^d$International Centre for Theoretical Physics Asia-Pacific, Beijing/Hangzhou, China \\
		${}^e$National Astronomical Observatories, Chinese Academy of Sciences, Beijing 100101, China    
\end{footnotesize}}

\begin{abstract}
The recently reported result from XENON1T experiment indicates that there is an excess with $3.5\sigma$ significance in the electronic recoil events. Interpreted as new physics, new sources are needed to account for the electronic scattering. We show that a dark fermion $\psi$ with mass about $ \mathcal{O}\left(10\right)$ MeV from semi-annihilation of $Z_3$ dark matter $X$ and subsequent decay of a dark gauge boson $V_\mu$ may be responsible for the excess. The relevant semi-annihilation process is $X+X\rightarrow \overline{X}+V_\mu (\rightarrow \psi + \overline{\psi})$, in which the final $\psi$ has a box-shape energy spectrum. The fast-moving $\psi$ can scatter with electron through an additional gauge boson that mixes with photon kinetically. The shape of the signals in this model can be consistent with the observed excess. The additional interaction with proton makes this model testable in future searches for nucleus recoil as well. Because of the lightness of invisible particle introduced, this scenario requires non-standard cosmology to accommodate the additional radiation component. 
\end{abstract}	
	
\maketitle

\section{Introduction}
Recently, XENON1T collaboration has reported an excess with $3.5\sigma$ significance in the electronic recoil events with an exposure of $0.65$ tonne-year~\cite{Aprile:2020tmw} data. The excess is observed around energy range $2-3~\keV$, which could be the relevant range for solar axion search~\footnote{This search also has some implications for neutrino magnetic moment, see Refs.~\cite{Aprile:2020tmw, Bell:2005kz, Bell:2006wi}.}. However, the interpretation of vanilla solar axion is in strong conflict with other astrophysical bounds on stellar cooling~\cite{Viaux:2013lha, Ayala:2014pea, Bertolami:2014wua, Giannotti:2017hny}. Alternative explanations are then needed for such an excess. 

Although the conservative scenario with tritium can not be excluded or confirmed at the moment, various relevant interpretations of new physics, constraints and connections have been explored timely in~\cite{Takahashi:2020bpq, Alonso-Alvarez:2020cdv, Nakayama:2020ikz, An:2020bxd, Kannike:2020agf, Fornal:2020npv, Du:2020ybt, Chen:2020gcl, Su:2020zny, Lee:2020wmh, Jho:2020sku, Bloch:2020uzh, Budnik:2020nwz, Harigaya:2020ckz, Bally:2020yid, Boehm:2020ltd, Khan:2020vaf, Primulando:2020rdk, Cao:2020bwd, Paz:2020pbc, AristizabalSierra:2020edu, Buch:2020mrg, Baryakhtar:2020rwy, Bramante:2020zos, Choi:2020udy, Bell:2020bes, Dey:2020sai, DiLuzio:2020jjp, Lindner:2020kko, Gao:2020wer, Zu:2020idx, McKeen:2020vpf, Smirnov:2020zwf, Chala:2020pbn, An:2020tcg, Baek:2020owl}. For instance, the excess could be connected to the absorption of bosonic dark matter (DM)~\cite{Takahashi:2020bpq, Alonso-Alvarez:2020cdv, Nakayama:2020ikz, An:2020bxd} (axion-like particle and dark photon), boosted DM~\cite{Kannike:2020agf, Fornal:2020npv, Du:2020ybt, Chen:2020gcl}, inelastic scattering~\cite{Su:2020zny, Lee:2020wmh, Jho:2020sku, An:2020tcg, Baek:2020owl}. 

In this study, we present an explanation of the excess in the framework of semi-annihilating DM with $Z_3$ symmetry in a viable microscopic model. The model was originally proposed and investigated in different contexts~\cite{Ko:2014nha, Ko:2014loa} by the present authors. Here, we show that in different parameter space, the semi-annihilation of DM with mass around $ \mathcal{O}\left(50\right)$ MeV can produce an unstable gauge boson that decays into a pair of dark fermions. The resulting boosted dark fermions can then interact with the electrons through dark photon with kinetic mixing and induce the electronic recoil signals. The event spectrum can be well consistent with the XENON1T observation. We also put an upper bound on the semi-annihilation cross section of DM and lower bound on the electron-scattering cross section. Due to the mixing with photon, dark fermions can also scatter with protons and produce nucleus recoil, distinguishable from other models with electronic signals only. 

This paper is organized as follows. In Sec.~\ref{sec:model} we present the model setup by introducing the particle contents and Lagrangian. Then in Sec.~\ref{sec:kin} we discuss the detailed kinematics that would be relevant for later investigations. Later, we give both analytic estimation and numerical illustration how the signal in this model can fit the XENON1T excess in Sec.~\ref{sec:event} and present constraints on the relevant cross sections in Sec~\ref{sec:const}. Finally, we summarize our paper.

\section{The Model}\label{sec:model}
The Lagrangian we are considering is the following one with 2 complex scalars $X$ and $\Phi$, 
a fermion $\psi$, and 
two dark $U(1)_D \times U(1)_D'$ gauge groups with two dark gauge bosons, 
$V_\mu$ for $U(1)_D$ and $A_\mu^{'}$ for $U(1)_D'$: 
\begin{align}
\mathcal{L}\supset &\left(D_\mu X\right)^\dagger D^\mu X + \left(D_\mu \Phi\right)^\dagger D^\mu \Phi - \lambda_X(\Phi^\dagger X^3 + \Phi X^{\dagger 3})-\overline{\psi}\left(i\gamma^\mu D_\mu -m_\psi\right)\psi \nonumber \\
&-\frac{1}{4}V_{\mu\nu}V^{\mu\nu}-\frac{1}{4}F_{\mu\nu}^{'}F^{'\mu\nu}-\frac{1}{2}m^2_{A'}A_\mu^{'}A^{'\mu}-\frac{\epsilon}{2}F_{\mu\nu}^{'}F^{\mu\nu}-V\left(X,\Phi,H\right),
\end{align}
The charge assignments of these fields are listed in TABLE ~\ref{default}.
Here the covariant derivatives are defined as 
\[D_\mu X=(\partial_\mu - igV_\mu)X,\ D_\mu \Phi = (\partial_\mu - i3g V_{\mu} )\Phi,\ D_{\mu}\psi= (\partial_\mu -ig Q_\psi V_\mu - if A'_\mu ) \psi ,\] 
with $g$ and $f$ are the gauge couplings for $U(1)_D$ and $U(1)_D^{'}$, respectively.  
Here we have assigned the $U(1)_D$ charges of $X$ and $\Phi$ to be $1$ and $3$, respectively. 
They are neutral under $U(1)_D'$ however.
The fermion $\psi$ may have a different $U(1)_D$ change $Q_\psi$, but should also be charged under $U(1)_D'$.  The $U(1)_D'$ gauge field $A'_\mu$ has the kinetic 
mixing with  ordinary photon $A_\mu$, and will induce $\psi$-electron 
scattering for XENON1T excess. The mass of $A'$ can originate from the usual Higgs mechanism or 
St\"{u}ckelberg trick, which does not affect our discussions in this paper. $A'$ may be connected 
to other hidden sector, which would relax its experimental constraints. Different implementations 
of $Z_3$ symmetry in other contexts have been investigated in~\cite{Belanger:2014bga, Aoki:2014cja, Bernal:2015bla, Choi:2015bya, Ma:2015mjd, Cai:2015tam, Ding:2016wbd, Hektor:2019ote, Kang:2017mkl, Athron:2018ipf, Kannike:2019mzk, Choi:2020ara}. 

\begin{table}[tb]
\caption{$U(1)_D \times U(1)_D'$ charge assignments of the fields}
\begin{center}
\begin{tabular}{|c|c|c|c|}
\hline
Fields & $~X~$ & $~\Phi ~$ & $~ \psi ~$ \\
 \hline 
$U(1)_D$ charges & $1$ & $3$ &  $Q_\psi$ \\
\hline
$U(1)_D'$ charges & $0$ & $0$ &  $ 1$ \\
\hline
\end{tabular}
\end{center}
\label{default}
\end{table}%

After the spontaneous symmetry breaking of $U(1)_D$, the scalar $\Phi$ has a non-zero vacuum 
expectation value $v_\phi$,  $\Phi \rightarrow (v_\phi+\phi)/\sqrt{2}$, and the gauge field $V_\mu$ gets its mass. In the scalar potential we also have the cubic term $(X^3+ X^{\dagger3})$ that preserves the discrete 
$Z_3$ symmetry, $X\rightarrow \exp(i2n\pi/3)X$, which makes $X$ a stable DM candidate but have the semi-annihilating process with emission of $V_\mu$ or (dark) Higgs (see Refs.~\cite{Ko:2014nha, Ko:2014loa} for more detail).

The masses and interaction terms of scalar fields are included collectively in the potential 
$V(X,\Phi, H)$, where $H$ is the Higgs doublet in standard model. Interaction with $H$ can make 
the dark sector in contact with thermal bath, which would be needed for thermal production of $X$. 
For non-thermal production, we do not have to specify the form and strength in $V(X,\Phi, H)$.

At one-loop level, $\psi$ would induce the kinetic mixing between $V$ and $A'$, whose mixing parameter would be at order of $\sim \dfrac{gf}{16\pi^2}$. As a result, $V$ can also mix with ordinary photon with parameter $\sim \dfrac{gf\epsilon}{16\pi^2}$. The physical effects due to these induced mixing will be discussed along the constraints that we shall explore in later sections. 

\section{Kinematics}\label{sec:kin}
\begin{figure}
	\includegraphics[width=0.4\textwidth,height=0.25\textwidth]{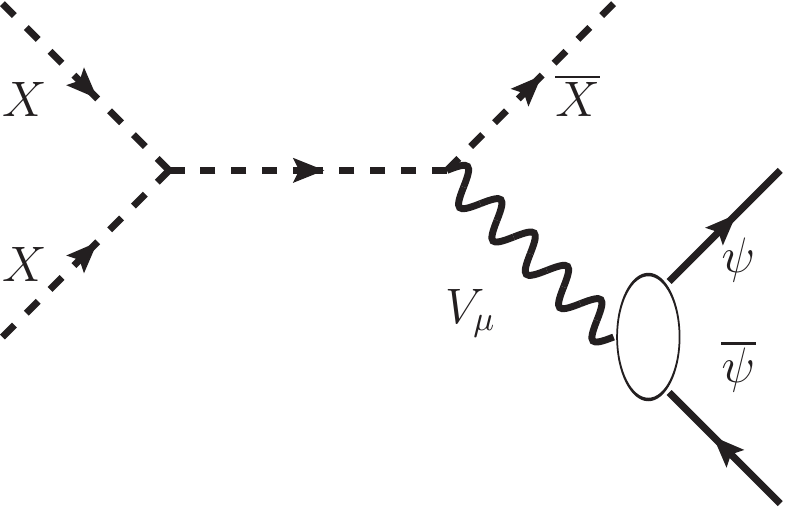}
	\caption{The typical Feynman diagram of semi-annihilation process, $ X+X\rightarrow \overline{X} + V_\mu$, with subsequent on-shell decay $V_\mu \rightarrow \psi + \bar{\psi}$. 
		\label{fig:semi}}
\end{figure} 
The relevant semi-annihilation process is shown in Fig.~\ref{fig:semi}, 
\begin{eqnarray}
 X+X\rightarrow \overline{X} + V_\mu(\rightarrow \psi \bar{\psi}),
\end{eqnarray}
where the DM $X$ with $Z_3$ symmetry semi-annihilates into its antiparticle and a dark photon $V_\mu$ which decays into dark $\psi$ pair subsequently. Choosing the mass differences properly, the resulting $\psi$ with velocity $v_\psi \sim 0.1$c is the boosted dark fermion that interacts with electron by an additional gauge interaction $A'_\mu $ that mixes photon through kinetic term. Here, we focus on the phenomenology in XENON1T. General discussions about boosted DM in other context can be found in~\cite{Agashe:2014yua, Kim:2019had, Aoki:2018gjf, Kopp:2015bfa, Kong:2014mia, Berger:2014sqa}. 

In principle, there is another annihilation process $ X+\overline{X}\rightarrow V_\mu^* \rightarrow \psi + \overline{\psi}$ where we have the corresponding energy and velocity,
\begin{equation}
E_\psi \simeq  m_X,\; \textrm{and }\; v_\psi = \sqrt{1-\frac{m_\psi^2}{m^2_X}}.
\end{equation}
Here and after, $m_i, i=X, V,\psi$ are the masses for particles, $X,V_\mu,\psi$, respectively. Simple estimation implies that we have $v_\psi \sim 0.1$c for $m_\psi \simeq 0.995m_X$, and $v_\psi \sim 0.87$c for $m_\psi \simeq 0.5m_X$. However, this process is velocity suppressed at the present time when DM in the Milky Way is moving with velocity $v\sim 10^{-3}$c, and the cross section would be $\sim 10^{-6}$ smaller than that in semi-annihilation. Therefore, we shall not consider this channel in our later discussions.

There are also contributions from annihilation $ X+\overline{X}\rightarrow V_\mu + V_\mu$, whose cross section depends on the gauge coupling $\sim g^4$, in comparison with that for the semi-annihilation $\sim \lambda_X^2$ (This relation is due to Goldstone equivalence theorem, which in this context states that the production of longitudinal mode of $V_\mu$ would be equivalent to the scalar $\phi$ when $m_X\gtrsim m_V$, namely, $X+X\rightarrow \overline{X}+\phi$). For simplicity, we shall focus on the parameter region $\lambda_X\gg g^2$ such that the semi-annihilation always dominates in our later discussions. Note that the opposite limit $\lambda_X\ll g^2$ would make $ X+\overline{X}\rightarrow V_\mu + V_\mu$ dominant, which also works with  the boosted $V_\mu$ that decays, although with a different kinematics~\footnote{ 
In Ref.~\cite{,Ko:2014loa}, this channel was exploited to explain galactic center $\gamma$-ray excess assuming  $m_V \lesssim m_X$.}. 

In the semi-annihilation we have the relations for the energies of final states,
\begin{equation}
E_V=\frac{3m^2_X+m^2_V}{4m_X}, E_X = \frac{5m^2_X-m^2_V}{4m_X}.
\end{equation}
The velocity of $V_\mu $ is
\begin{equation}
v_V =\frac{1}{3m^2_X+m^2_V} \sqrt{\left[4m^2_X-\left(m_{X}+m_V\right)^{2}\right]\left[4m^2_X-\left(m_{X}-m_V\right)^{2}\right]}.
\end{equation} 
The final $\psi$ particles have an energy distribution with box shape, 
\begin{equation}
2=\int_{E_-}^{E_+}dE  \frac{dN}{dE}, \;\frac{dN}{dE}=\frac{2}{E_+-E_-}\mathcal{\theta}(E_-,E_+),
\end{equation}
where $\mathcal{\theta}(E_-,E_+)=1$ for $E_-<E<E_+ $ and zero otherwise, $E_{\pm}=E_V\left(1\pm\beta_V \beta_\psi ^{\ast}\right)/2$, $\beta _V = v_V$ and $\beta_\psi ^{\ast} = \sqrt{1-4m^2_\psi/m_V^2}$. It can be also translated into velocity distribution $f_\psi$,
\begin{equation}
2=\int_{v_-}^{v_+}dv_\psi  f_\psi ,\;f_\psi = \frac{2m_\psi v_\psi}{\left(E_+-E_-\right)\left(1-v^2_\psi\right)^{3/2}},
\end{equation}
where $v_\psi = \sqrt{1-m_\psi^2/E^2_\psi}$. The energy interval of the distribution is $E_V\beta_V \beta_\psi ^{\ast}$, which depends on the three masses, and the relative half-width is $\delta=\beta_V \beta_\psi ^{\ast}$. For small mass differences, $m_X-m_V\ll m_V$ and/or $m_V/2-m_\psi \ll m_\psi$, the spectrum will be very narrow around $E_V/2$. 

\section{Event Rate}\label{sec:event}
The fast-moving $\psi$ can scatter with electron through $A'_\mu$ interaction and induce prompt scintillation events (S1) at XENON1T experiment. The maximal recoil energy of electron is $E_R\simeq 2m_e v_\psi ^2$ for $m_\psi \gg m_e$ (throughout our later estimations, we neglect the dependence on the scattering angle, a valid approximation due to the sharp detector efficiency drop below $\sim 2$~keV). Since the events of excess are centered around $E_R\sim 2.5~\keV$, we would need $v_\psi \simeq 0.05$c. The differential rate of such events can be estimated as
\begin{equation}
\frac{d R}{d E} = n_T   \langle \Phi_\psi \sigma_{e}(E) \rangle,
\end{equation}
where $n_T\sim 4.6\times 10^{27}/$ton, $\Phi_\psi$ is the flux of $\psi$ and $\sigma_{e}$ is the scattering cross section between $\psi$ and electron. To explain the excess, we would need $dR/dE\sim 30/$(ton yr keV), which gives 
\begin{equation}
\langle \Phi_\psi \sigma_e \rangle \simeq 2.4\times 10^{-35}/(\textrm{s keV}).
\end{equation}
The $\langle \cdot \rangle$ denotes that we shall take into account the smearing effect due to energy resolution and efficiency of the experiments~\cite{Aprile:2020tmw}. The energy resolution~\cite{Aprile:2020yad} is parametrized by Gaussian distribution with uncertainty $\sigma$,
\begin{equation}
\frac{\sigma}{E} =\left(\frac{31.71}{\sqrt{E/\keV}}+0.15\right)\%.
\end{equation}
For the reconstructed energy at $E\simeq 2.7~\keV$, the relative resolution is about $19.45\%$. All of these effects are included in our later numerical illustrations. 

The flux $\Phi_\psi$ from annihilation of DM of our galaxy is given by
\begin{align}
	\Phi_\psi &= \frac{v_\psi}{4\pi}\frac{dN}{dE}\frac{\langle \sigma_{\textrm{ann}} v \rangle}{4}\int d\Omega \int ds \frac{\rho_X^2}{m_X^2} \nonumber \\
	&\simeq  2\times 10^{-5}~\textrm{cm}^{-2}\textrm{s}^{-1}\times \left(\frac{\GeV}{m_X}\right)^2\left(\frac{\langle \sigma_{\textrm{ann}} v \rangle}{6\times 10^{-26}~\textrm{cm}^{3}\textrm{/s}}\right)\left(\frac{v_\psi}{0.05\textrm{c}}\right)\frac{dN}{dE},
\end{align}
where the $\rho_X$ is the energy density of $X$ with an NFW profile, ${dN}/{dE}$ is the number distribution of $\psi$ at production and the integration is performed over the line-of-sight $s$ and all-sky solid angle $\Omega$. 

The above analytic estimation shows that the $\psi$-$e$ scattering cross section should be around 
\begin{equation}\label{eq:csflux}
\sigma_e \sim 5\times 10^{-31}~\textrm{cm}^2 \times \left(\frac{m_X}{\GeV}\right)^2\left(\frac{6\times 10^{-26}~\textrm{cm}^{3}\textrm{/s}}{\langle \sigma_{\textrm{ann}} v \rangle}\right)\left(\frac{0.05\textrm{c}}{v_\psi}\right),
\end{equation}
with canonically thermal annihilation cross section. Increasing the mass of DM would decrease the flux of $\psi$ and correspondingly requires larger scattering cross section $\sigma_{e}$ for compensation of the flux loss. As we shall show later, $\sigma_{e}$ should be less than $6\times 10^{-29}~\textrm{cm}^2$, otherwise the boosted $\psi$ would be loosing too much energy or even stopped by the earth, unable to reach the underground detector.

\begin{figure}
	\includegraphics[width=0.73\textwidth,height=0.55\textwidth]{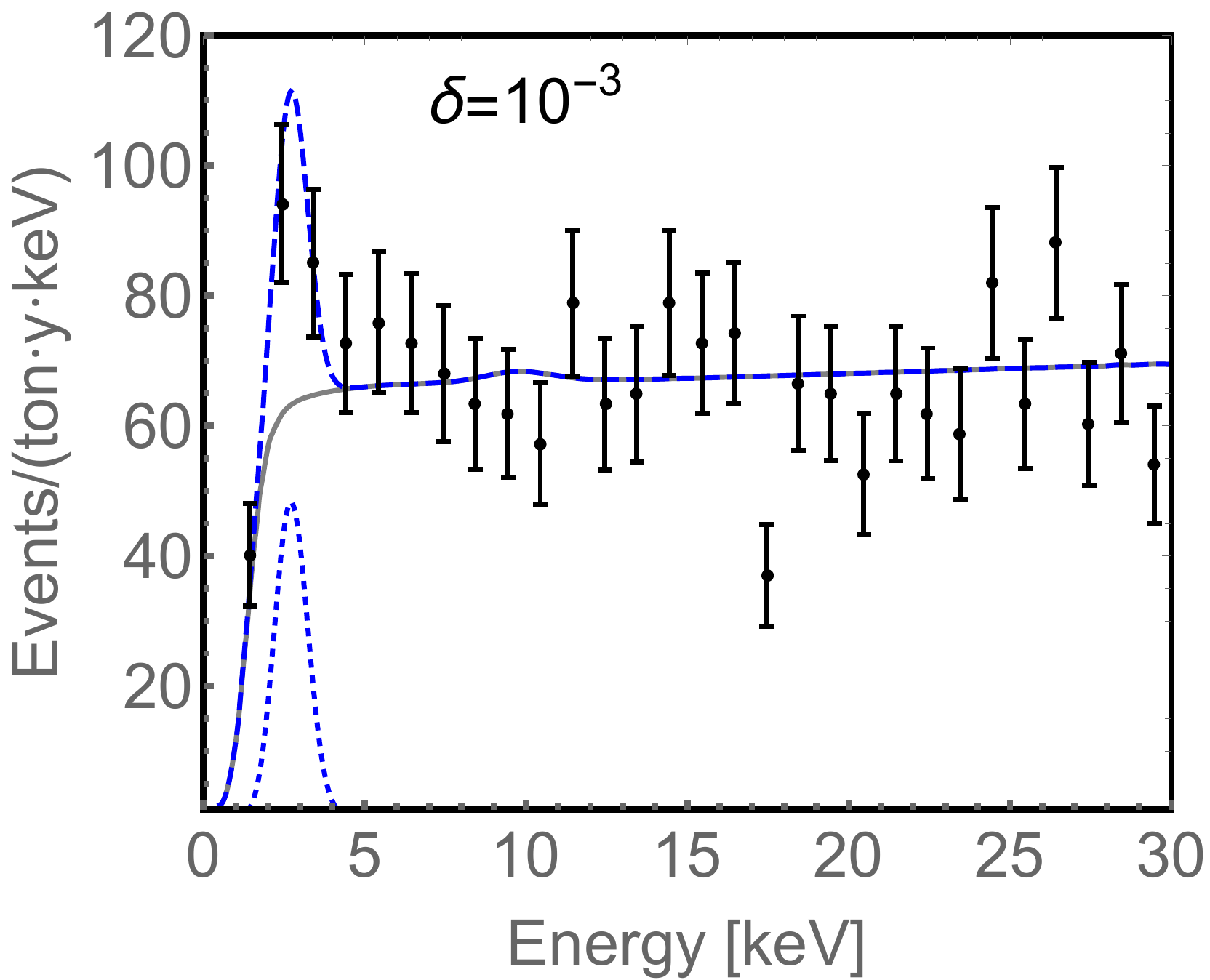}
	\caption{The illustration of signal shapes with box-shape spectra for boosted $\psi$ from $V_\mu$'s decay. The relative half-width is taken as $\delta =  0.1\%$, shown as the dotted curves. The black dots with error bars and the gray background curved are extracted from XENON1T results. The central value of the recoil energy is chosen at $E_R\simeq 2.7~\keV$. 
		\label{fig:excess}}
\end{figure} 

In Fig.~\ref{fig:excess} we illustrate schematically with a box-shape energy spectrum having relative half-width, $\delta = 0.1\%$. The central value of recoil energy is chosen at $E_R\simeq 2.7~\keV$. The black dots with error bars and the gray background curve are extracted from XENON1T result~\cite{Aprile:2020tmw}. The dotted curve describes the signal shapes with $\delta = 0.1\%$ and the dashed curve is the sum of background and signal. Apparently, the signal is consistent with the reported excess. The value of $\delta$ then can be directly translated into the requirements on the mass 
differences through the relation $\delta = \beta_V\beta_\psi^\ast$. Together with $v_\psi\simeq 0.05$c, 
we can determine the two mass ratios of $m_V/m_X$ and $m_\psi/m_V$. 

\begin{figure}
	\includegraphics[width=0.7\textwidth,height=0.55\textwidth]{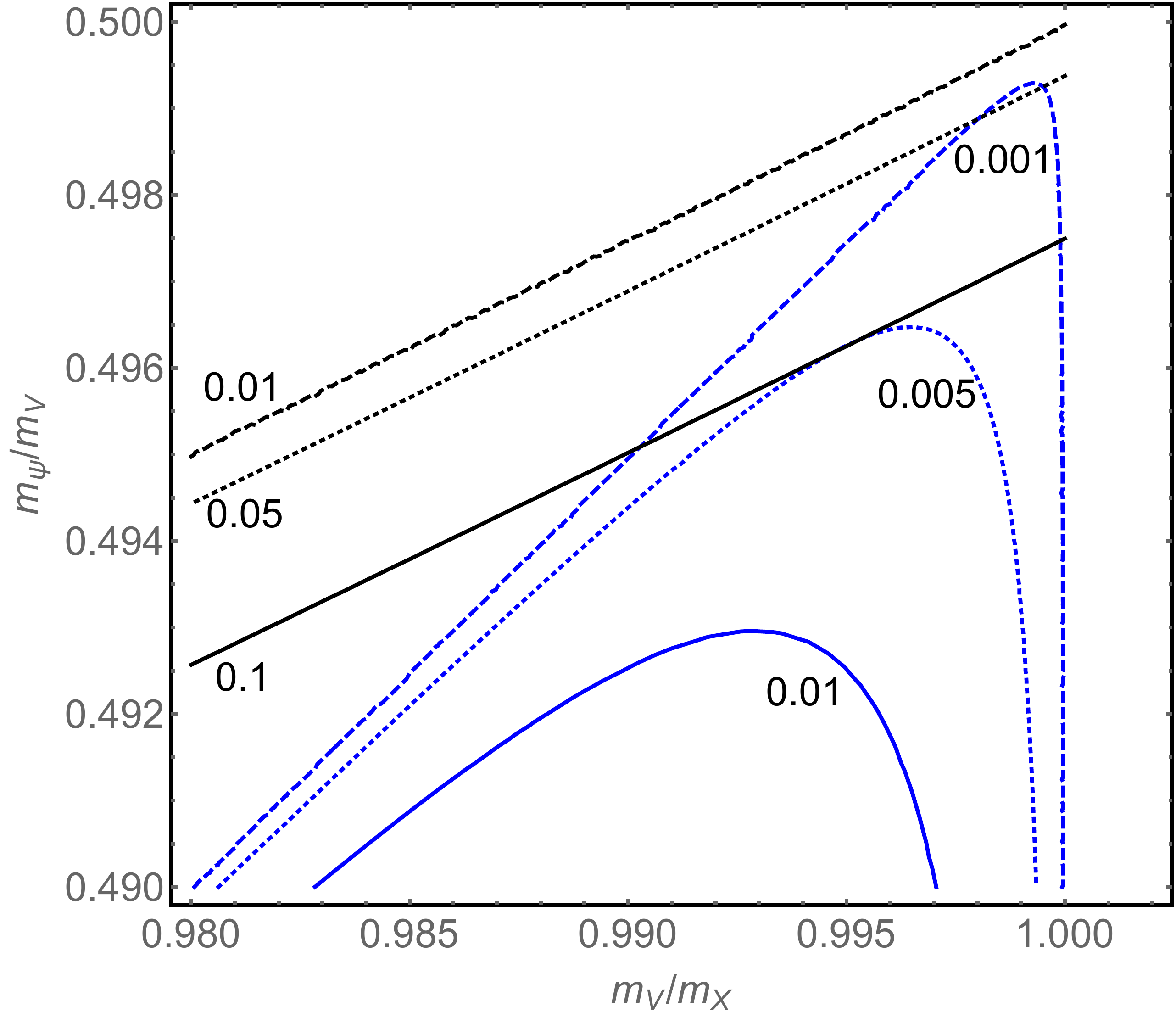}
	\caption{The three straight lines correspond to the velocity of $\psi$, $v_\psi=0.01$c (dash), 0.05c (dot), 0.1c (solid). The three blue curves are the contours for relative half-width $\delta = 0.001$ (dash), 0.005 (dot), 0.01 (solid).
		\label{fig:ratio}}
\end{figure}

To precisely determine the favored parameter region, dedicated statistical analysis would be needed and is beyond the scope of this paper. Instead, we only show several contours for the relative ratios of $m_V/m_X$ and $m_\psi/m_V$ in Fig.~\ref{fig:ratio} to provide the proof of concept. The three straight lines in the figure correspond to the different velocities of $v_\psi=0.01,0.05,0.1$c (from top to down) and the numbers near the three blue curves indicates the relative half-width 
$\delta=0.1\%,0.5\%,1\%$ (from top to down), respectively. The choice of $v_\psi\simeq 0.05$c 
would suggest $\delta \simeq 0.1\%$, $m_V/m_X\simeq 0.998$ and $m_\psi/m_V\simeq 0.499$. And more conservative values of $v_\psi\simeq 0.1$c can enlarge the available parameter space significantly and make $\delta\sim 0.5\%$ viable.

\section{Constraints}\label{sec:const}
\subsection{Relic Density}
In this section, we shall show that it is possible to produce DM $X$ through semi-annihilation with the correct relic abundance. The semi-annihilation cross section is estimated as
\begin{equation}
\sigma_{\textrm{ann}}v\simeq \frac{\lambda_X^2}{4\pi m_X^2}ps(s),\; ps(s)\equiv\sqrt{1-\left(m_X+m_V\right)^2/s},
\end{equation}
where $\sqrt{s}$ is the total energy in the centre-of-mass reference frame and $ps(s)$ is the factor from phase space. The typical velocity of DM particle is about $0.3$c at the time of freeze-out for thermally-produced DM, and is $\sim 10^{-3}$c in Milky Way. For $m_V/m_X=0.998$, we can calculate the ratio of $ps(s)$, 
\begin{equation}
\frac{ps(s_\textrm{fo})}{ps(s_\textrm{mw})}\simeq 7.2.
\end{equation}
To give the correct relic abundance $\Omega_X\simeq 0.25$, we shall have $\langle \sigma_{\textrm{ann}}v\rangle_\textrm{fo}\simeq 6\times 10^{-26}~\textrm{cm}^{3}\textrm{/s}$ at freeze-out, which determines the mass-coupling relation and the current value of annihilation cross section,
\begin{equation}
m_X/\lambda_X\simeq 2200~\GeV,\;\langle \sigma_{\textrm{ann}}v\rangle_\textrm{now}\simeq 0.83\times 10^{-26}~\textrm{cm}^{3}\textrm{/s}. 
\end{equation}
The above estimations demonstrate that $\lambda_X\simeq 4\times 10^{-4}$ would be sufficient for $\GeV$ DM, which is definitely viable in this model setup since all the annihilating products are in dark sector.    

\subsection{Galactic Bound}
As we have demonstrated in previous sections, the event rate depends on the product of the DM semi-annihilation cross section and $\psi$-$e$ scattering cross section, $\langle \sigma_{\textrm{ann}} v \rangle \sigma_e$. Larger $\langle \sigma_{\textrm{ann}} v \rangle$ implies smaller $\sigma_e$. There is some degeneracy in these two values, which is perfectly fine since it means the viable parameter space can be vast. 

However, there are upper bounds on $\langle \sigma_{\textrm{ann}} v \rangle$ from astrophysics even if all the final annihilation products are in dark sector. Because the resulting $X$ from semi-annihilation is also fast moving and may escape from the dark halo in our galaxy, we should require the annihilation rate is smaller than 1 per Hubble time, then we obtain the upper bound,
\begin{equation}\label{eq:upper}
\frac{\langle \sigma_{\textrm{ann}} v \rangle}{m_X}\lesssim 3 \times 10^{-18}\;\frac{\textrm{cm}^3/\textrm{s}}{\GeV}.
\end{equation}
When deriving the above result, we have used the averaged DM density within $r<1~$kpc $\overline{\rho}_X\simeq 1~\GeV/$cm$^3$ with the generalized NFW profile~\cite{Navarro:1995iw}, parametrized as   
\begin{equation}\label{eq:haloprofile}
	\rho\left(r\right)=\rho_\odot \left[\frac{r_\odot}{r}\right]^\gamma \left[\frac{1+r_\odot/r_c}{1+r/r_c}\right]^{3-\gamma},
\end{equation}
where $r_c\simeq 20$~kpc, $\gamma=1.26$ and $\rho_\odot \simeq 0.3~\textrm{GeV}/\textrm{cm}^3$ around solar system.

We should note that if $\langle \sigma_{\textrm{ann}} v \rangle \gg 6\times 10^{-26}~\textrm{cm}^3/\textrm{s}$, $X$ can not be produced thermally in the early universe since its relic density would be too small. In such a case, other production mechanism would be needed. For instance, $X$ could be produced from other heavy particles' decay or some more involved cosmology with hidden sectors.

If the cross section of self-scattering, $X+X\rightarrow X+X$, is too large, DM semi-annihilation might lead to core formation because the annihilation product $X$ can be absorbed by the DM particles in the galactic center~\cite{Chu:2018nki, Kamada:2019wjo}, which is illustrated with the following typical annihilation and self-interacting cross sections, 
\begin{equation}
\langle \sigma_{\textrm{ann}} v \rangle \simeq 6 \times 10^{-26} ~\textrm{cm}^3/\textrm{s},\;
\sigma_X/m_X \simeq 0.1~\textrm{cm}^2/\textrm{g}. 
\end{equation}
The effect of core formation depends on the product of the above two quantities and is more considerable in smaller halos~\cite{Chu:2018nki, Kamada:2019wjo}. In the scenario with thermal cross section, both $ \sigma_{\textrm{ann}} $ and $\sigma_X$ are at weak-scale. In such a case we may neglect the core formation. However, as we reach the limit of Eq.~\ref{eq:upper}, core formation would be an interesting feature of this model. Note that in the dedicated studies~\cite{Chu:2018nki, Kamada:2019wjo} the available kinetic energy after semi-annihilation is about $m_X/4$, while here in our interested parameter spaces it is about $0.002 m_X$. Though depending on the parameter, the corresponding limit is comparable to the one in Eq.~(\ref{eq:upper})~\footnote{We thank the referee to point out this to us.}.

\subsection{Direct Detection}
In order to explain the XENON1T excess, the above upper bound eq.~(\ref{eq:upper}) correspondingly gives a lower bound on $\sigma_e$,
\begin{equation}\label{eq:lower}
\sigma_e\gtrsim 10^{-38}~\textrm{cm}^2\times \frac{m_X}{\GeV}.
\end{equation}
This limit can be easily satisfied for $\GeV$-scale DM as we shall show below. Note that the constraints from cosmic-ray scattering with DM~\cite{Bringmann:2018cvk,Ema:2018bih} do not apply here because the density of $\psi$ particles in the galactic background is much lower than DM $X$. If the Higgs-portal and $V_\mu$ kinetic portal couplings are small, then the interaction between $X$ and standard model particles would also be suppressed, therefore the scattering between $X$ and cosmic rays is rare and the constraints can be relaxed, but potentially detectable in future.

The above requirement of $\sigma_e$ can be satisfied with a viable kinetic mixing parameter 
$\epsilon$ for $\gamma-A'$ and dark photon mass $m_{A'}$, 
\begin{equation}
\sigma_e \simeq 4\pi \epsilon^2 \frac{\alpha \alpha' m^2_e }{m^4_{A'}}\simeq 1.0\times 10^{-33}~\textrm{cm}^2\left(\frac{\epsilon^2}{10^{-6}}\right)\left(\frac{\alpha'}{10^{-2}}\right) 
\left(\frac{10~ \MeV}{m_{A'}}\right)^4.
\end{equation} 
Here $\alpha = 1/137$ is the fine-structure constant and $\alpha'=f^2/4\pi$ is the constant for $A'$. Such a light $A'$ can be searched in various experiments. For $A'$ that mainly decays into invisible particles, the experimental constraints~\cite{Essig:2013lka} allow $\epsilon\lesssim  10^{-4}$ for $\MeV$-scale $A'$ and $\epsilon\lesssim  10^{-3}$ at $\mathcal{O}(10)$ $\MeV$, respectively. This cross section can also be consistent with Eq.~(\ref{eq:csflux}), for example, with parameter set, $m_{A'}\sim 10~\MeV, m_X\simeq 50~\MeV, \epsilon\simeq 10^{-3}$ and $\alpha'\simeq 0.03$ with thermal annihilation cross section. 

Since $A'$ mixes with photon, it can also mediate the interaction between $\psi$ and proton. The scattering cross section for sub-GeV $\psi$ with proton is given by
\begin{equation}
\sigma_p \simeq 4\pi \epsilon^2 \frac{\alpha \alpha' m^2_\psi }{m^4_{A'}}=\frac{m^2_\psi}{m^2_e}\sigma_e.
\end{equation}

The fermion $\psi$ is also a potential DM candidate, although its relic density might be significantly smaller than that of $X$. The annihilation cross section of $\psi + \bar{\psi}\rightarrow A'+A'$ at freeze-out can be calculated as
\begin{equation}
	\sigma v \sim \frac{4\pi \alpha'^2}{m^2_\psi}\approx \frac{4f^4}{\lambda^2_X}\frac{ \lambda_X^2}{4\pi m^2_X}, \; \alpha'\equiv f^2/4\pi,
\end{equation}
where in the last step we have used the mass relation $m_\psi\approx m_X/2$. It is evidently that the relic density of $\psi$ is suppressed by the factor $\lambda^2_X/(4 f^4)$, in comparison with that of $X$. For $\GeV$-scale $\psi$ and $f\simeq 1$, its relic density would be $\Omega_\psi\sim 10^{-8}\Omega_X$. As long as $f^2\gg \lambda_X/2$, $\psi$'s contribution to DM relic density would be sub-dominant.

The DM direct search experiments can put constraints on the scattering rate $R_{p/e}$ between $\psi$ and proton or electron,
\begin{equation}
	R_{p/e}\propto \frac{\Omega_\psi \sigma_{p/e}}{m_\psi}.
\end{equation}
For sub-GeV DM, the most stringent constraint on $\sigma_p$ is from {\textit{CRESST-III 2019}}~\cite{Abdelhameed:2019hmk}, which is about $\sigma_p\lesssim 10^{-37}~$cm$^2$ for $m_\psi=1~\GeV$, and $\sigma_p\lesssim 10^{-34}~$cm$^2$ for $m_\psi=0.1~\GeV$, assuming $\Omega_\psi=0.25$. The limit is getting weaker as $m_\psi$ decreases because the recoil energy of the proton will be below the detection threshold. Then DM-electron scattering would prevail. For instance, {\textit{DarkSide}}~\cite{Agnes:2018oej} gives $\sigma_{e}\lesssim 10^{-38}~$cm$^2$ for $m_\psi=0.1~\GeV$ if $\Omega_\psi=0.25$. Since we expect $\Omega_\psi$ is not the dominant DM component, the above constraints are much relaxed for $\Omega_\psi\ll \Omega_X$.

The boosted $\psi$ from semi-annihilation can also scatter with proton whose recoil energy is around 
\begin{equation}
E_R\sim \frac{\left(m_\psi v_\psi\right)^2}{2m_N}\simeq 3~\keV \left(\frac{m_\psi}{0.5~\GeV}\right)^2 
\left(\frac{v_\psi}{0.05c}\right)^2 \left(\frac{100~\GeV}{m_N}\right).
\end{equation}
Since the recoil energy is around keV-scale, the signal would be similar to an ordinary heavy DM particle's scattering. However, the flux of the boosted $\psi$ is $\sim 10^{-10}$ smaller than that of $\sim 30~\GeV$ DM. We referred to $30~\GeV$ DM because its direct search bound is the strongest one, $\sigma_p\lesssim 5\times 10^{-47}~$cm$^2$~\cite{Aprile:2018dbl}. Taking the flux into account, we get $\sigma_p\lesssim 5\times 10^{-37}~$cm$^2$ for the boosted $\psi$. Again, we emphasize that the limit would become much weaker as $m_\psi$ is getting smaller. Because the recoil energy would be below the detection threshold when $m_\psi \lesssim 0.1~$GeV. The above argument also indicates that future experiments may be able to probe the boosted $\psi$ through nucleus recoils, in addition to the electron recoil signals. 

\begin{figure}
	\includegraphics[width=0.5\textwidth,height=0.35\textwidth]{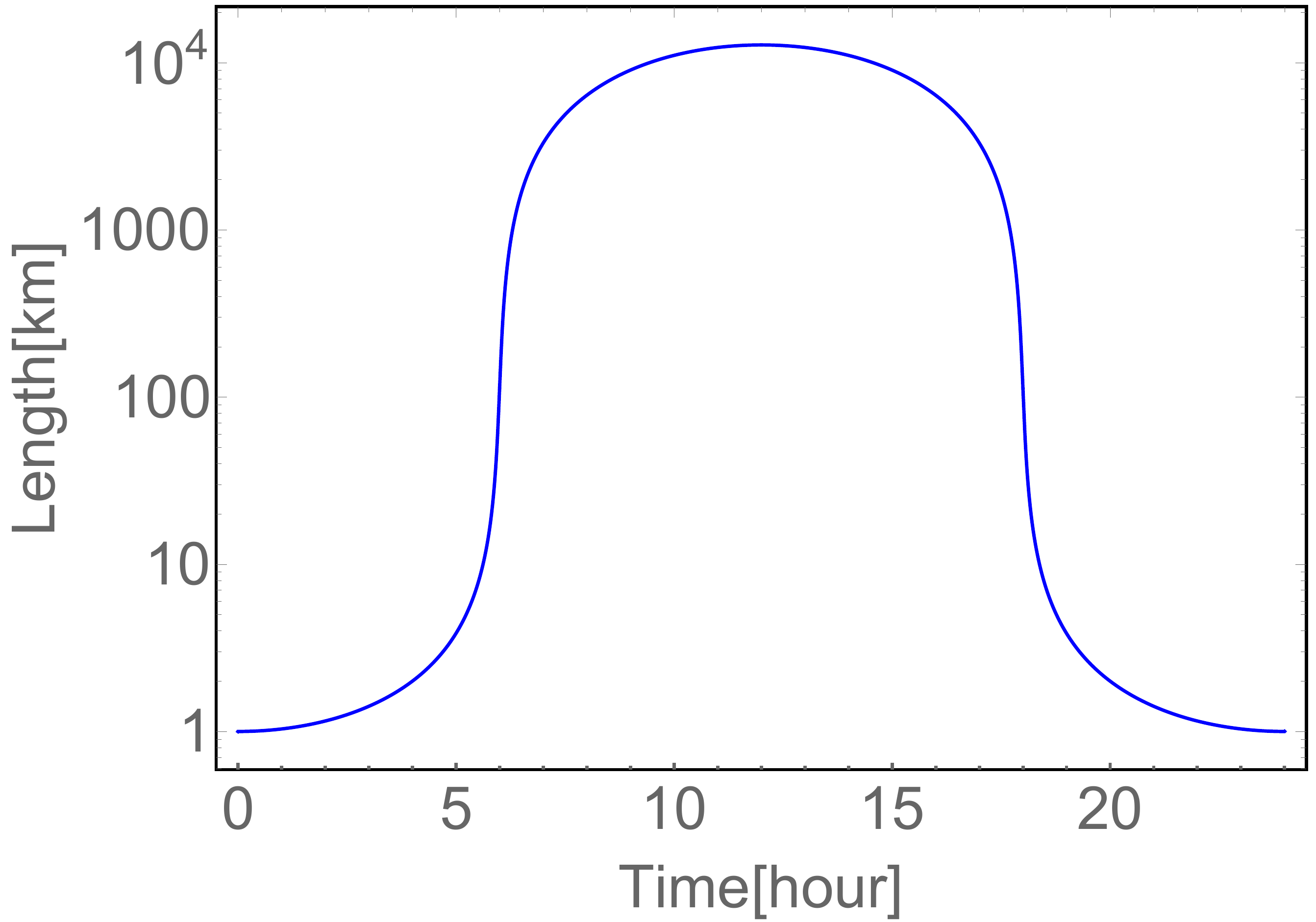}
	\caption{A schematic plot of the traveled length in earth of $\psi$ as a function of time in $24$ hours. The minimal length is set to $1$~km for illustration. 
		\label{fig:modulation}}
\end{figure}

The interaction between $\psi$ with electron or proton may also modulate the flux of boosted $\psi$ if the scattering rate is large enough~\cite{Emken:2019tni}. The free-streaming path of $\psi$ in earth can be estimated as
\begin{equation}
L_{\textrm{fs}}\sim 60~\textrm{km}\times \left(\frac{10^{-31}~\textrm{cm}^2}{\sigma_{p/e}}\right),
\end{equation}
where we have used the mean density of earth, $\rho \simeq 5.5~$g$/$cm$^3$. For $L_{\textrm{fs}}\gtrsim 1$~km for typical underground DM direct search experiment, we shall have $\sigma_{p/e}\lesssim 6\times 10^{-30}~$cm$^2$. When combined with Eq.~(\ref{eq:csflux}), this leads to an upper bound for the mass $m_X$, 
\begin{align}
m_X&\simeq 4~\GeV\times\left(\frac{\sigma_{e}}{6\times 10^{-30}~\textrm{cm}^2}\right)^{1/2}
\left(\frac{6\times 10^{-26}~\textrm{cm}^{3}\textrm{/s}}{\langle \sigma_{\textrm{ann}} v \rangle}\right)^{-1/2}\left(\frac{0.05\textrm{c}}{v_\psi}\right)^{-1/2}\\
&\simeq 60~\MeV\times\frac{30~\MeV}{m_\psi}\left(\frac{\sigma_{p}}{6\times 10^{-30}~\textrm{cm}^2}\right)^{1/2}
\left(\frac{6\times 10^{-26}~\textrm{cm}^{3}\textrm{/s}}{\langle \sigma_{\textrm{ann}} v \rangle}\right)^{-1/2}\left(\frac{0.05\textrm{c}}{v_\psi}\right)^{-1/2}.
\end{align}
In the last step, we have used the relation, $\sigma_p =\frac{m^2_\psi}{m^2_e}\sigma_e$, for $\psi$'s interaction. Note that for $L_{\textrm{fs}}\simeq 1$~km, only those $\psi$s that traveled a length no much larger that $L_{\textrm{fs}}$ can reach the detector energetically. Therefore, there would be a daily modulation in the signals due to the rotation of the earth. In Fig.~\ref{fig:modulation}, we illustrate schematically how the traveled distance in earth depends on time in a day. The minimal length is set to 1~km at time $t=0$ and radius of earth $R\simeq 6387$~km is used. The signal rate would be smaller at the time domain with longer length. We note that the plot should only be understood schematically, not literally, because the detailed calculation would involve more information about the XENON1T experiment site, which is beyond the scope of the paper.

\subsection{$N_\textrm{eff}$}

In the above discussion, we have assumed that gauge boson $A'$ dominantly decays into some invisible light particle, $\varphi$. This setup is subject to constraints on the effective number of neutrinos $N_\textrm{eff}$ if the light particle $\varphi$ was in thermal equilibrium before the big-bang nucleosynthesis (BBN). In the standard $\Lambda$CDM paradigm (cold DM (CDM) with cosmological constant $\Lambda$)~\footnote{Six parameters are needed to specify the  paradigm, $\Omega_b h^2, \Omega _c h^2, \theta_{\textrm{MC}},\tau, A_s$ and $n_s$. See Ref.~\cite{Planck:2018} for detailed definition.}, current limit on the extra relativistic species is $\delta N_\textrm{eff}\lesssim 0.3-0.5$~\cite{Planck:2018} (the precise value depends on the data sets chosen for the analysis), which seems severely constraining and disfavors the scenario with thermal light particle. However, we should be reminded that the constraints on $\delta N_\textrm{eff}$ highly depend on the cosmological models. For example, the analysis~\cite{Kreisch:2019yzn} in the framework of self-interacting dark radiation demonstrates that $\delta N_\textrm{eff}=0.98\pm 0.29$ is preferred to address the Hubble tension.

Note that although $A'$ may be in equilibrium with the thermal bath, it does not pose a problem for $\delta N_\textrm{eff}$. Because $\mathcal{O}\left(10\right)$-MeV $A'$ can essentially transfer its energy into $e^{\pm}$ and $\gamma$ before BBN through the fast process, $A'+e^{\pm}\rightarrow e^{\pm} + \gamma $, there is no additional contribution to $N_\textrm{eff}$ from $A'$. Requiring $A'$ decouples before $e^{\pm}$ do, we shall have $m_\psi \gtrsim \mathcal{O}\left(10\right)$ MeV and $m_X\simeq m_V \gtrsim \mathcal{O}\left(20\right)$ MeV.

\subsection{Indirect Detection}

Because of the induced mixing between $V$ and photon $A$, standard model fermion pairs can be produced from the decay of $V$, which is from the semi-annihilation $X+X\rightarrow \overline{X} + V$. Then, DM indirect searches from cosmic-ray, gamma-ray and cosmic microwave background (CMB) experiment would constrain the effective annihilation cross section, 
\begin{equation}
\langle \sigma v \rangle_\textrm{eff}=\langle \sigma_\textrm{ann} v \rangle\times \mathcal{B}r,
\end{equation}
where $\mathcal{B}r$ is the branch ratio of $V$'s decay into standard model particles, $\mathcal{B}r\simeq 10^{-4}\times \alpha'\epsilon^2$. Even if we take the large possible values, $\alpha'=1,\epsilon=10^{-4}$ and $\langle \sigma_\textrm{ann} v \rangle = 3\times 10^{-18}~$cm$^3\textrm{/s}$, $\langle \sigma v \rangle_\textrm{eff}$ is smaller than the upper bound~\cite{Leane:2018kjk} from CMB, $4.1\times 10^{-28}~$cm$^3\textrm{/s}\times m_X/\GeV$. Limits from cosmic-ray and gamma-ray are less stringent than that from CMB in the sub-GeV range of our interest. Therefore, the model setup can satisfy the limits from indirect searches.

\section{Summary}\label{sec:summary}
In this paper, we have presented an explanation of the recently observed excess in electronic recoil events at XENON1T experiment with a viable microscopic dark matter (DM) model with local $Z_3$ discrete symmetry. The semi-annihilation of DM $X$ can generate an unstable gauge boson $V_\mu $ which subsequently decays into dark fermion pair $\psi+\overline{\psi}$. The energy spectrum of dark fermion has a box shape, with the width depending on the relative mass differences. Because of the semi-annihilation and decay, $X+X\rightarrow \overline{X}+V_\mu (\rightarrow \psi + \overline{\psi})$, the final dark fermion can be fast-moving with velocity around $0.05$c. Then it scatters with an electron through the dark photon that has kinetic mixing with ordinary photon field. We have shown the resulting spectrum shape can be consistent with the observed data by XENON1T with viable parameter values with DM mass around $\mathcal{O}(50)$~MeV. The correlation between the scattering of boosted $\psi$ with electrons and protons distinguishes this model from other ones, which usually introduce the interaction with electrons solely. The lightness of the invisible particles is confronted with $N_\textrm{eff}$ constraints, which would require non-standard cosmology. 

\begin{Large}
\textbf {\newline Acknowledgments}
\end{Large}$\\$
We would like to thank the anonymous referee for helpful suggestions on various improvements of our discussions. The work of P.K. is supported in part by KIAS Individual Grant (Grant No. PG021403) at Korea Institute for Advanced Study and by National Research Foundation of Korea (NRF) Grant No. NRF-2019R1A2C3005009, funded by the Korea government (MSIT). YT is supported by National Natural Science Foundation of China (NSFC) under Grants No.~11851302 and by the Fundamental Research Funds for the Central Universities. 



%

\end{document}